\documentclass[a4paper,11pt]{article}
\usepackage{makeidx}
\usepackage[english]{babel}
\usepackage[latin1]{inputenc}
\usepackage{amsfonts}
\usepackage{hyperref}
\usepackage{bm}
\usepackage{amssymb}
\usepackage{latexsym}
\usepackage{amsmath}
\usepackage{amscd}
\usepackage{amsthm}
\usepackage{mathrsfs}
\usepackage{amsmath,amstext}
\usepackage{graphicx}
\usepackage{color}
\usepackage{fancyhdr}
\usepackage{epsfig}
\usepackage{alltt}

\theoremstyle{plain}

\theoremstyle{definition}

\title{HIV with contact-tracing: a case study in Approximate Bayesian Computation}

\author{Michael G.B. Blum$^1$, Viet Chi Tran$^2$\\
{\footnotesize $^1$ CNRS, Laboratoire TIMC-IMAG, Universit\'e Joseph Fourier, Grenoble, France}\\
{\footnotesize $^2$ Laboratoire P. Painlev\'e, Universit\'e Lille 1, France}
}

\date{}
\numberwithin{equation}{section}

\newcommand{\E}{\mathbb{E}}

\newcommand{\R}{\mathbb{R}}

\newcommand{\etal}{\textit{et al. }}
\newcommand{\eg}{\textit{e.g. }}

\begin{document}

\maketitle

\begin{abstract}
Missing data is a recurrent issue in epidemiology where the infection process may be partially observed. Approximate Bayesian Computation, an alternative to data imputation methods such as Markov Chain Monte Carlo integration, is proposed for making inference in epidemiological models. It is a likelihood-free method that relies exclusively on numerical simulations. ABC consists in computing a distance between simulated and observed summary statistics and weighting the simulations according to this distance. We propose an original extension of ABC to path-valued summary statistics, corresponding to the cumulated number of detections as a function of time. For a standard compartmental model with Suceptible, Infectious and
Recovered individuals (SIR), we show that the posterior distributions obtained with ABC and MCMC
are similar. In a refined SIR model well-suited to the HIV contact-tracing data in Cuba, we perform a comparison between ABC with full and binned detection times. For the Cuban data, we evaluate the efficiency of the detection system and predict the evolution of the HIV-AIDS disease. In particular, the percentage of undetected infectious individuals is found to be of the order of $40\%$.
\end{abstract}


\section{Introduction}

Mathematical modelling plays an important role for understanding and predicting the spread of diseases, as well as for comparing and evaluating public health policies. Although deterministic modelling can be a guide for describing epidemics, stochastic models have their importance in featuring realistic processes and in quantifying confidence in parameters estimates and prediction uncertainty \cite{Beckerbritton}. Standard models in epidemiology consist in compartmental models in
which the population is
structured in different classes composed of Susceptible, Infectious,
and Recovered (or Removed) individuals (SIR). Parameter estimation for SIR models is usually a difficult task because of missing observations, which is a recurrent issue in epidemiology. When the infected population is partially observed or when the infection times are missing, the computation of the likelihood is numerically infeasible because it involves integration over all the unobserved events.

Markov Chain Monte Carlo (MCMC) methods that treat the missing data as extra parameters, have become increasingly popular for calibrating stochastic epidemiological models with missing data \cite{oneillroberts,oneill,cauchemcarrat}. However, MCMC may be computationally prohibitive for high-dimensional missing observations \cite{cauchemezferguson, chisstersinghferguson} and fine tuning of the proposal distribution is required for efficient algorithms \cite{GilksRoberts}. In this paper, we show that SIR models with missing observations can be calibrated with Approximate Bayesian Computation (ABC), an alternative to MCMC, originally proposed for making inference in population genetics \cite{beaumontzhangbalding}. This approach is not based on the likelihood function but relies on numerical simulations and comparisons between simulated and observed summary statistics.

This work is motivated by the study of the Cuban HIV-AIDS database that contains the dates of detection of the 8,662 individuals that have been found to be HIV positive in Cuba between 1986 and 2007 \cite{auvert}. The database contains additional covariates including the manner by which an individual has been found to be HIV positive. The individuals can be detected either by {\it random screening} (individuals `spontaneously' take a detection test) or {\it contact-tracing}. Contact-tracing consists in testing the sexual contacts of detected individuals \cite{hsieh}. The total number of infectious individuals as well as the infection times are unknown.

In Section \ref{sectionSIR}, we introduce the stochastic SIR model with contact-tracing developed by \cite{arazozaclemencontran}. Section \ref{sectionABC} is devoted to ABC methods when all detection times are known. We propose an original extension of ABC to path-valued summary statistics consisting of the cumulated number of detections through time. For a simple SIR model, we compare numerically the posterior distributions obtained with ABC and MCMC.
Section 4 deals with possibly noisy or binned detection times for which the previous path-valued statistics are unavailable. We introduce a finite dimensional vector of summary statistics and compare the statistical properties of point estimates and credibility intervals obtained with full and binned detection times. Finally, Section \ref{sectioncuba} concentrates on the analysis of the Cuban HIV-AIDS database. We address several questions concerning the dynamic of this epidemic: what is the percentage of the epidemic that is known \cite{hsieh,arazozacubana,auvert}; how many new cases of HIV are expected in the forthcoming years; and what is the proportion of detections that is expected in the contact-tracing program.

\section{A stochastic SIR model for HIV-AIDS epidemics with contact-tracing}\label{sectionSIR}
\begin{figure}[h]
\begin{center}
\includegraphics[height = 5cm]{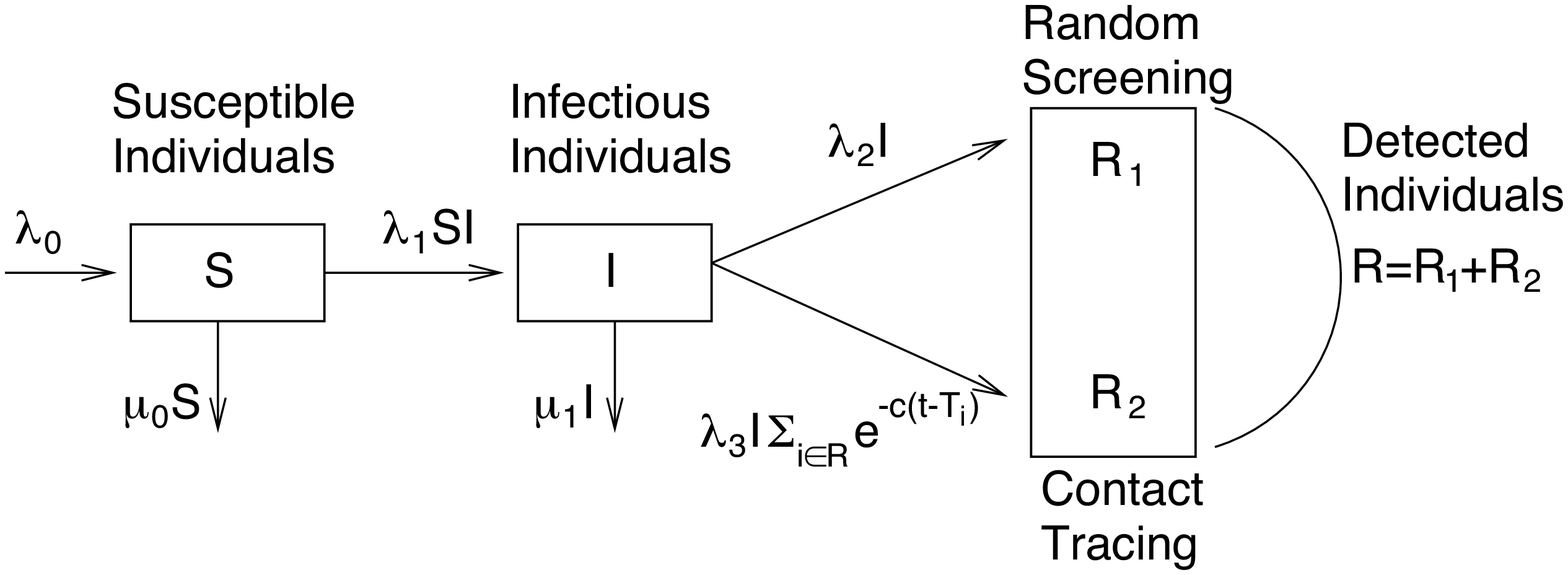}
\end{center}
\caption{Schematic description of the SIR model with contact-tracing.}
\label{fig:sir}
\end{figure}

We restrict our study to the sexually-transmitted epidemic of HIV in Cuba (90\% of the epidemic, see \cite{hsieh}). For modelling the dynamics of the number of known and unknown HIV cases, we consider a SIR model developed by \cite{arazozaclemencontran}. The population is divided into three main classes $S$, $I$ and $R$ corresponding to the {\it susceptible}, {\it infectious}, and \textit{detected} individuals considered as {\it removed} because we assume that they do not transmit HIV anymore (see Figure \ref{fig:sir}). The population of the susceptible individuals, of size $S_t$, at time $t>0$, consists of the sexually active seronegative individuals. Individuals immigrate into the class $S$ with a rate $\lambda_0$ and leave it by dying/emigrating, with rate $\mu_0 S_t$, or by becoming infected. The class of infectious individuals, of size $I_t$, corresponds to the seropositive individuals who have not taken a detection test yet and may thus contaminate new susceptible individuals. We assume that each infected individual may transmit the disease to a susceptible individual at rate $\lambda_1$ so that the total rate of infection is equal to $\lambda_1 S_t I_t$. Individuals leave the class $I$ when they die/emigrate with a total rate of $\mu_1 I_t$, or when they are detected to be HIV positive.

The class $R$ of the detected individuals, of size $R_t$, is subdivided into two subclasses. We denote by $R^1_t$ (resp. $R^2_t$) the size of the removed population detected by random screening (resp. contact-tracing). The total rate of detection by random screening is $\lambda_2 I_t$. For the rate of contact-tracing detection, the model shall capture the fact that the contribution of a removed individual depends on the time elapsed since she/he has been found to be HIV positive. We consider the two following expressions for the total rate of contact-tracing detection
\begin{equation}
\lambda_3 I_t \sum_{i \in {R}} e^{-c(t-T_i)}\quad \mbox{ and }\quad \lambda_3 I_t \sum_{i \in {R}} e^{-c(t-T_i)}/(I_t+ \sum_{i \in {R}} e^{-c(t-T_i)}),
\label{formlambda3}
\end{equation}
where $T_i$ denotes the time at which a removed individual $i$ has been detected. The weight $\exp(-c(t-T_i))$, with $c>0$, determines the contribution of a removed individual $i$ to the contact-tracing according to the time $t-T_i$ she/he has been detected. The first rate in (\ref{formlambda3}) corresponds to a mass action principle, and is proportional to the sum of the contributions of the detected individuals. The second rate in (\ref{formlambda3}) corresponds to a model with frequency dependence. In the following, $\theta=(\mu_1,\lambda_1,\lambda_2,\lambda_3,c)$ denotes the multivariate parameter of the model. The parameters of interest are here $\lambda_1$, $\lambda_2$ and $\lambda_3$.

\subsection{Connection between the stochastic and the deterministic SIR model}\label{sectionpde}

Here we consider the first rate of contact-tracing detection in (\ref{formlambda3}), but similar results can be obtained for the second rate. \cite{arazozaclemencontran} showed that in a large population limit, the SIR process of Section \ref{sectionSIR} can be modelled with stochastic differential equations and converges to the solution of the following system of ordinary differential equations
\begin{equation}\label{eqgdepop}
\left \{
\begin{array}{lcl}
\frac{ds_t}{dt} & = & \lambda_0  - \mu_0 s_t-\lambda_1 s_t i_t \\
\frac{di_t}{dt} & = &  \lambda_1 s_t i_t -(\mu_1+\lambda_2)i_t - \lambda_3 i_t r_t\\
\frac{dr_t}{dt}& = & \lambda_2 i_t +\lambda_3 i_t r_t -c r_t
\end{array}\right.,
\end{equation}
where $s_t$ and $i_t$ denote the size of the susceptible and infectious populations, and where $r_t$ is the contribution of the removed individuals to the contact-tracing (see Section 1 of the supplementary material for more details).

Apart from the inherent stochastic nature of epidemic propagation that may be particularly important for small populations \cite{finken}, considering a stochastic SIR model rather than its deterministic counterpart can present at least two important advantages. First, it is quite straightforward to perform simulations from the stochastic model (see Section 2 of the supplementary material) and this is one motivation for considering ABC methods.
 Second, the individual-centered stochastic process suits the formalism of statistical methods, which are based on samples of individual data. Since the estimates of the stochastic process converge to the parameters of the ODEs \cite{arazozaclemencontran}, ABC provides a new alternative for calibrating parameters of ODEs \cite{banksburnscliff,burnscliffdoughty}.

\section{ABC with sufficient summary statistics for epidemic models}\label{sectionABC}

\subsection{Main principles of ABC}\label{sectionmainpple}

For simplicity, we deal here with densities and not general probability measures. Let $\mathbf{x}$ be the available data and $\pi(\theta)$ be the prior. Two approximations are at the core of ABC.\\

\noindent \textbf{Replacing observations with summary statistics} Instead of focusing on the posterior density $p(\theta\, |\, \mathbf{x})$, ABC aims at a possibly less informative \textit{target} density $p(\theta\, |\, S(\mathbf{x})=s_{obs})\propto {\textrm Pr}(s_ {obs} |\theta)\pi(\theta)$ where $S$ is a summary statistic that takes its values in a normed space, and $s_{obs}$ denotes the observed summary statistic. The summary statistic $S$ can be a $d$-dimensional vector or an infinite-dimensional variable such as a $L^1$ function. Of course, if $S$ is sufficient, then the two conditional densities are the same. The target distribution will also be coined as the \textit{partial posterior distribution}.\\

\noindent \textbf{Simulation-based approximations of the posterior} Once the summary statistics have been chosen, the second approximation arises when estimating the partial posterior density  $p(\theta\, |\, S(\mathbf{x})=s_{obs})$ and sampling from this distribution. This step involves nonparametric kernel estimation and possibly correction refinements given in Section \ref{sec:correction}.

\subsection{Sampling from the posterior}\label{sectionsamplingposterior}

The ABC method with smooth rejection generates random draws from the target distribution as follows \cite{beaumontzhangbalding}
\begin{enumerate}
\item Generate $N$ random draws $(\theta_i,s_i)$, $i=1, \dots , N$. The parameter $\theta_i$ is generated from the prior distribution $\pi$ and the vector of summary statistics $s_i$ is calculated for the $i^{th}$ data set that is simulated from the generative model with parameter $\theta_i$.
\item Associate to the $i^{th}$ simulation the weight $W_i=K_{\delta}(s_i-s_{obs})$, where $\delta$ is a tolerance threshold and $K_\delta$ a (possibly multivariate) smoothing kernel.
\item The distribution $(\sum_{i=1}^N W_i \mathbf{\delta}_{\theta_i})/(\sum_{i=1}^N W_i)$, in which $\mathbf{\delta}_{\theta}$ denotes the Dirac mass at $\theta$, approximates the target distribution. \end{enumerate}

\subsection{Point estimation and credibility intervals}\label{Point estimation}

Once a sample from the target distribution has been obtained, several estimators may be considered for point estimation of each one-dimensional parameter $\lambda_j$, $j=1,2,3$. Using the weighted sample $(\lambda_{j,i},W_i)$, $i=1,\dots ,N$, the \textit{mean} of the target distribution $p(\lambda_j|s_{obs})$ is estimated by
    \begin{equation}
    \hat{\lambda_j}=\frac{\sum_{i=1}^N \lambda_{j,i} W_i}{\sum_{i=1}^N W_i}=\frac{\sum_{i=1}^N \lambda_{j,i} K_\delta(s_i-s_{obs})}{\sum_{i=1}^N K_\delta(s_i-s_{obs})}, \quad j=1,2,3 \label{estimateurbayesien}
    \end{equation}
which is the well-known Nadaraya-Watson regression estimator of the conditional expectation $\E(\lambda_j\, |\, s_{obs})$. We also compute the \textit{medians}, \textit{modes}, and $95\%$ credibility intervals (CI) of the marginal posterior  distribution (see Section 3 of the supplementary material).

\subsection{Data and choice of summary statistics}\label{sectionsummarystatistic}

\textbf{Data} Starting at the time of the first detection in 1986, the Cuban HIV-AIDS data consist principally of the detection times at which the individuals have been found to be HIV positive. At the time of the last detection event, in July 2007, there is a total of 8,662 individuals in the database. For each detection event, there is a label indicating if the individual has been detected by random screening or contact-tracing.\\

\noindent \textbf{Summary statistics} We consider the two (infinite-dimensional) statistics $(R^1_t,t\in [0,T])$ and $(R^2_t,t\in [0,T])$. Their sum is equal to the cumulated number of detections since the beginning of the epidemic. Because the data consist of the detection times for the two different types of detection, these two statistics can simply be viewed as a particular coding of the whole dataset so that the partial posterior distribution $p(\theta\, |\, R^1,R^2)$ is equal to the posterior distribution $p(\theta\, |\, \mathbf{x})$.

The $L^1$-norm between the $i^{\rm{th}}$ simulated path $R^l_i$ and the observed one $R^l_{obs}$ is
\begin{equation}
 \label{eqn:l1norm}
\|R^l_{obs}-R^l_i\|_1=\int_0^T  |R^l_{obs,s}- R^l_{i,s}|\, ds \quad , \; l=1,2 , \; i=1, \dots , N.
\end{equation}
 For computing the weights $W_i$, we choose a product kernel so that $W_i=K_{\delta_1}(\|R^1_{obs}-R^1_i\|_1) K_{\delta_2}(\|R^2_{obs}-R^2_i\|_1)$ where $\delta_1$, $\delta_2$ are 2 tolerance thresholds. Epanechnikov kernels are considered for $K_{\delta_1}$ and $K_{\delta_2}$ and $\delta_1$ and $\delta_2$ are found by accepting a given percentage $P_{\delta_1}=P_{\delta_2}$ of the simulations.

\subsection{Comparisons between ABC and MCMC methods for a standard SIR model}\label{sectionMCMC}

Following \cite{beaumontzhangbalding} a performance indicator for ABC techniques consists in their ability to replicate likelihood-based results given by MCMC. Here the situation is particularly favorable for comparing the two methods since the partial and the full posterior are the same. In the following examples, we choose samples of small sizes ($n=3$ and $n=29$) so that the dimension of the missing data is reasonable and MCMC achieves fast convergence. For large sample sizes with high-dimensional missing data, MCMC convergence might indeed be a serious issue and more thorough updating scheme shall be implemented \cite{cauchemezferguson, chisstersinghferguson}.

We consider the standard SIR model with no contact-tracing ($\lambda_3=0$). We choose gamma distributions for the priors of $\lambda_1$ and $\lambda_2$ with a shape parameters of $0.1$ and rate parameters of $1$ and $0.1$. The data consist of the detection times and we assume that the infection times are not observed. We implement the MCMC algorithm of \cite{oneillroberts}. A total of $10,000$ steps are considered for MCMC with an initial burn-in of $5,000$ steps. For ABC, the summary statistic consists of the cumulative number of detections as a function of time. A total of $100,000$ simulations are performed for ABC.

\begin{figure}[h]
\begin{center}
\includegraphics[height = 12cm,angle=270]{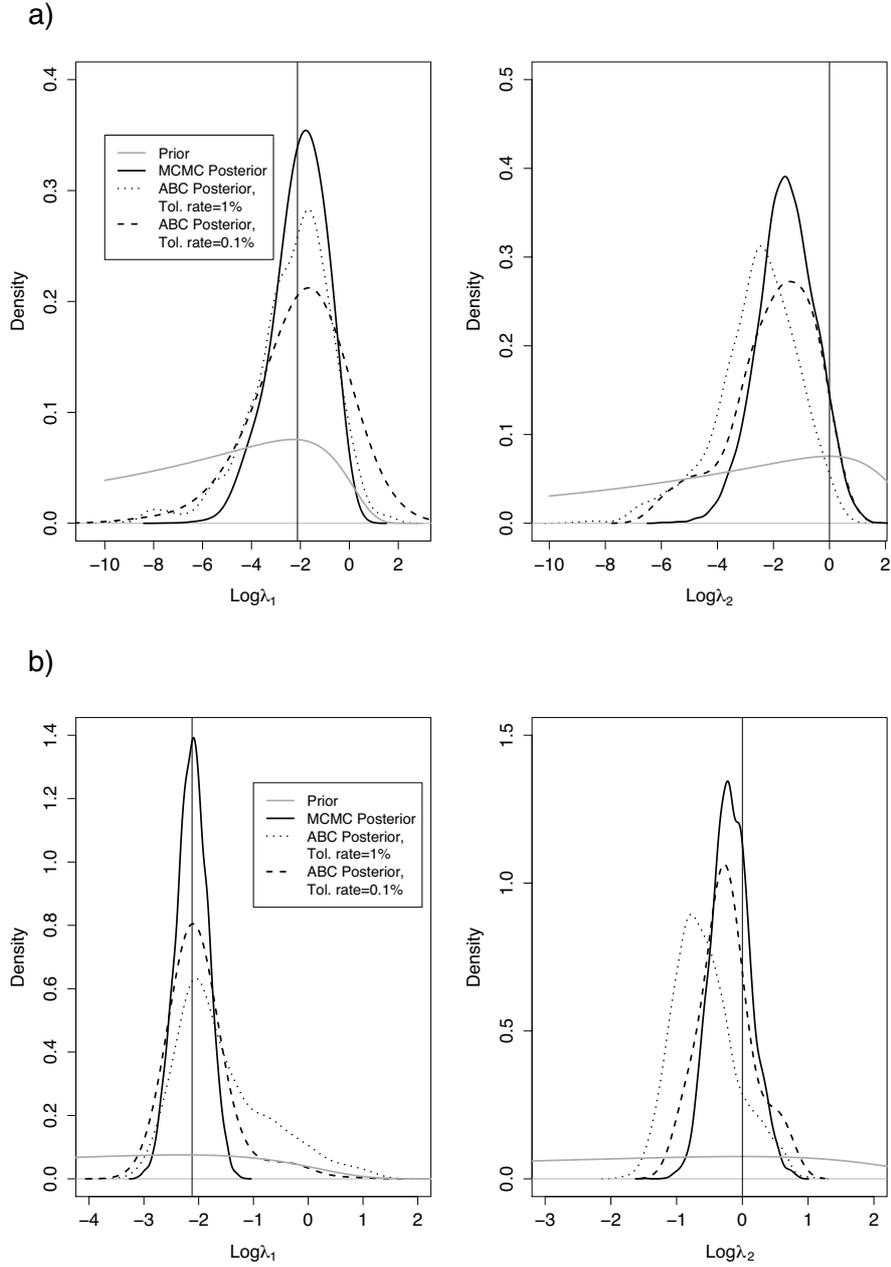}
\end{center}
\caption{Comparison of the posterior densities obtained with MCMC and ABC. The vertical lines correspond to the values of the parameters used for generating the synthetic data. a) The data consist of 3 detection times that have been simulated by \cite{oneillroberts}. b) The data consist of 29 detection times that we  simulated by setting $\lambda_1=0.12$, $\lambda_2=1$, $S_0=30$, $I_0=1$, and $T=5$ (see the supplementary material for the 29 detection times).}
\label{fig:3obs}
\end{figure}

The first example was previously considered by \cite{oneillroberts}. They simulated detection times by considering one initial infectious individual and by setting $S_0=9$, $\lambda_1=0.12$, and $\lambda_2=1$ (see Section 4 of the supplementary material for the data). As displayed by Figure \ref{fig:3obs}, the posterior distributions obtained with ABC are extremely close to the ones obtained with MCMC provided that the tolerance rate is sufficiently small. We see that the tolerance rate changes importantly the posterior distribution obtained with ABC (see the posterior distributions for $\lambda_2$).

In a second example, we simulate a standard SIR trajectory with $\lambda_1=0.12$, $\lambda_2=1$, $S_0=30$ and $I_0=1$. The data now consist of 29 detection times (see Section 4 of the supplementary material). Once again, Figure \ref{fig:3obs} shows that the ABC and MCMC posteriors are close provided that the tolerance rate is small enough. ABC produces posterior distributions with larger tails compared to MCMC, even with the lowest tolerance rate of $0.1\%$. This can be explained by considering the extreme scenario in which the tolerance threshold $\delta$ goes to infinity: every simulation has a weight of 1 so that ABC targets the prior instead of the posterior. As the prior has typically larger tails than the posterior, ABC inflates the posterior tails.

\section{Comparison between ABC with full and binned detection times}\label{section_vrai_ABC_SIR}

When there is noise or when the detection times have been binned, the full observations $(R^1_t,t\in [0,T])$ and $(R^2_t,t\in [0,T])$ are unavailable. Then, we replace these summary statistics by a vector of summary statistics such as the numbers of detections per year during the observation period. Since these new summary statistics are not sufficient anymore, the new partial posterior distribution may be different from the posterior $p(\theta \,|\,\mathbf{x})$. In the following, we compare point estimates and CIs obtained from ABC with full (method of Section \ref{sectionABC}) and binned detection times (method of Section \ref{sec:correction}).

\subsection{A new set of summary statistics}

We consider a d-dimensional vector of summary statistics of three different types. First, we compute the numbers $R^1_T$ and $R^2_T$ of individuals detected by random screening and contact-tracing by the end of the observation period. Second, for each year $j$, we compute the numbers of individuals that have been found to be HIV positive $R^l_{j+1}-R^l_j$, $l=1,2$. The third type of summary statistics consists of the numbers of new infectious for each of the the sixth first years $I_{j+1}-I_j$ for $j=0,\dots,5$, as well as the mean time during which an individual is infected but has not been detected yet. This mean time corresponds to the mean sojourn time in the class I for the sixth first years. These summary statistics are available here because infectious times before 1992 are known.

In order to compute the weights $W_i$, we consider the following spherical kernel $K_{\delta}(x)\propto K(\|\mathbf{H}^{-1}x\|/\delta)$. Here $K$ denotes the one-dimensional Epanechnikov kernel, $\|.\|$ is the Euclidian norm of $\R^d$ and $\mathbf{H}^{-1}$ a matrix. Because the summary statistics may span different scales, $\mathbf{H}$ is taken equal to the diagonal matrix with the standard deviation of each one-dimensional summary statistic on the diagonal.

\subsection{Curse of dimensionality and regression adjustments}\label{sec:correction}

In the case of a $d$-dimensional vector of summary statistics, the estimator of the conditional mean (\ref{estimateurbayesien}) is convergent if the tolerance rate satisfies $\lim_{N\rightarrow +\infty}\delta_N=0$, so that its bias converges to 0, and $\lim_{N\rightarrow +\infty}N\delta_N^d=+\infty$, so that its variance converges to 0 \cite{fan1}. As $d$ increases, a larger tolerance threshold shall be chosen to keep the variance small. As a consequence, the bias may increase with the number of summary statistics. This phenomenon known as the {\it curse of dimensionality} may be an issue for the ABC-rejection approach. The following paragraph presents regression-based adjustments that cope with the curse of dimensionality.

The adjustment principle is presented in a general setting within which the corrections of \cite{beaumontzhangbalding} and \cite{blumfrancois} can be derived. Correction adjustments aim at obtaining from a random couple $(\theta_i,s_i)$ a random variable distributed according to $p(\theta\,|\,s_{obs})$. The idea is to construct a coupling between the distributions $p(\theta|s_i)$ and $p(\theta | s_{obs})$, through which we can shrink the $\theta_i$'s to a sample of i.i.d. draws from $p(\theta | s_{obs})$. In the remaining of this subsection, we describe how to perform the corrections for each of the one-dimensional components separately. For $\theta\in \R$, correction adjustments are obtained by assuming a relationship $\theta=G(s,\varepsilon)=:G_s(\varepsilon)$ between the parameter and the summary statistics. Here $G$ is a (possibly complicated) function and $\varepsilon$ is a random variable with a distribution that does not depend on $s$. A possibility is to choose $G_s=F_s^{-1}$, the (generalized) inverse of the cumulative distribution function of $p(\theta|s)$. In this case, $\varepsilon=F_s(\theta)$ is a uniform random variable on $[0,1]$. The formula for adjustment is given by
\begin{equation}
\label{eqn:correc}
\theta_i^*=G_{s_{obs}}^{-1}(G_{s_i}(\theta_i)) \quad i=1, \dots , N.
\end{equation}For $G_s=F_s^{-1}$, the fact that the $\theta_i^*$'s are i.i.d. with density $p(\theta|s_{obs})$ arises from the standard inversion algorithm. Of course, the function $G$ shall be approximated in practice. As a consequence, the adjusted simulations $\theta_i^*$, $i=1,\dots ,N$, constitute an approximate sample of $p(\theta\,|\, s_{obs})$. The ABC algorithm with regression adjustment can be described as follows
\begin{enumerate}
\item Simulate, as in the rejection algorithm, a sample $(\theta_i,s_i)$, $i= 1,\dots, N$.
\item By making use of the sample of the $(\theta_i,s_i)$'s weighted by the $W_i$'s, approximate the function $G$ such that $\theta_i=G(s_i,\varepsilon_i)$ in the vicinity of $s_{obs}$.
\item Replace the $\theta_i$'s by the adjusted $\theta_i^*$'s. The resulting weighted sample $(\theta_i^{*},W_i)$, $i=1, \dots, N$, form a sample from the target distribution.
\end{enumerate}

\noindent \textbf{Beaumont \etal local linear regressions (LOCL)} The case where $G$ is approximated by a linear model $G(s,\varepsilon)=\alpha+ s^{t}\beta+\varepsilon$, was considered by \cite{beaumontzhangbalding}. The parameters $\alpha$ and $\beta$ are inferred by minimizing the weighted squared error $\sum_{i=1}^N K_{\delta}(s_i-s_{obs}) (\theta_i-(\alpha+(s_i-s_{obs})^{T}\beta))^2$.
Using (\ref{eqn:correc}), the correction of \cite{beaumontzhangbalding} is derived as
\begin{equation}
\label{eqn:adj}
\theta_i^{*}=\theta_i-(s_i-s_{obs})^{T}\hat{\beta},\; i=1, \dots, N.
\end{equation}
Asymptotic consistency of the estimators of the partial posterior distribution with the correction (\ref{eqn:adj}) is obtained by \cite{blum}.\\

\noindent \textbf{Blum and Fran\c{c}ois' nonlinear conditional heteroscedastic regressions (NCH)}
To relax the assumptions of homoscedasticity and linearity inherent to local linear regression, \cite{blumfrancois} approximated $G$ by $G(s,\varepsilon)=m(s)+\sigma(s) \times \varepsilon$
where $m(s)$ denotes the conditional expectation, and $\sigma^2(s)$ the conditional variance. The estimators $\hat{m}$ and $\log\hat{\sigma}^2$ are found by adjusting two feed-forward neural networks using a regularized weighted squared error. For the NCH model, parameter adjustment is performed as follows
$$
\theta_i^{*}=\hat{m}(s_{obs})+(\theta_i-\hat{m}(s_{i}))\times \frac{\hat{\sigma}(s_{obs})}{\hat{\sigma}(s_i)},\; i=1, \dots , N.
$$
In practical applications of the NCH model, we train $L=10$ neural networks for each conditional regression (expectation and variance) and we average the results of the $L$ neural networks to provide the estimates $\hat{m}$ and $\log\hat{\sigma}^2$.\\

\noindent \textbf{Reparameterization} In both regression adjustment approaches, the regressions can be performed on transformations of the responses $\theta_i$ rather that on the responses themselves. Parameters whose prior distributions have finite supports are transformed via the logit function and non-negative parameters are transformed via the logarithm function. These transformations guarantee that the $\theta_i^{*}$'s lie in the support of the prior distribution and have the additional advantage of stabilizing the variance.

\subsection{A comparison for simulated datasets}\label{sectionvalidation}

In order to work on data similar to the Cuban database, we simulate $M=200$ synthetic data sets  for the HIV-AIDS epidemic with $\mu_1=2\times 10^{-6}$, $\lambda_1=1.14\times 10^{-7}$, $\lambda_2=3.75\times 10^{-1}$, $\lambda_3=6.55\times 10^{-5}$, and $c=1$ \cite{arazozaclemencontran}.
The initial conditions are set to $S_0=6\times 10^{6}$, the size of the Cuban population in the age-group 15-49, $I_0=232$  and $R_0=0$ \cite{auvert}. Here we simulate only 6 years of the epidemics.

We study four variants of ABC for estimating $\lambda_1$, $\lambda_2$, and $\lambda_3$: one with the two path-valued summary statistics and three with the vector of summary statistics. When using the finite dimensional vector of summary statistics, we perform the smooth rejection approach as well as the LOCL and NCH corrections with a total of 21 summary statistics: the 18 summary statistics corresponding to the yearly increases of $R^1$, $R^2$, and $I$; the final numbers of detected individuals $R^1_6$ and $R^2_6$; and the mean sojourn time in the class $I$. Each of the $M=200$ estimations of the partial posterior distributions are performed using a total of $N=5000$ simulations of the SIR model with the mass action principle (first rate in (\ref{formlambda3})).\\

\noindent\textbf{Prior distributions} The prior distributions for $\mu_1, \lambda_1, \lambda_2$ and $\lambda_3$ are chosen to be uniform on a log scale. The choice of a log scale reflects our uncertainty about the order of magnitude of the parameters. The prior distribution for $\log_{10}(\mu_1)$ is $\mathcal{U}(-6,-4)$ where $\mathcal{U}(a,b)$ denotes the uniform distribution on the interval $(a,b)$. The prior distribution is $\mathcal{U}(-9,-6)$ for $\log_{10}(\lambda_1)$, $\mathcal{U}(-4,3)$ for $\log_{10}(\lambda_2)$, and $\mathcal{U}(-8,2)$ for $\log_{10}(\lambda_3)$. The bounds of the uniform distributions are set to keep the simulations from being degenerate. The prior for $c$ is $\log(2)/\mathcal{U}(1/12,5)$ so that the half-life of $t\mapsto e^{-ct}$, which measures the individual contribution to the detection by contact-tracing, is uniformly distributed between $1/12$ and $5$ years.\\

\begin{figure}[h]
\begin{center}
\includegraphics[height = 12cm]{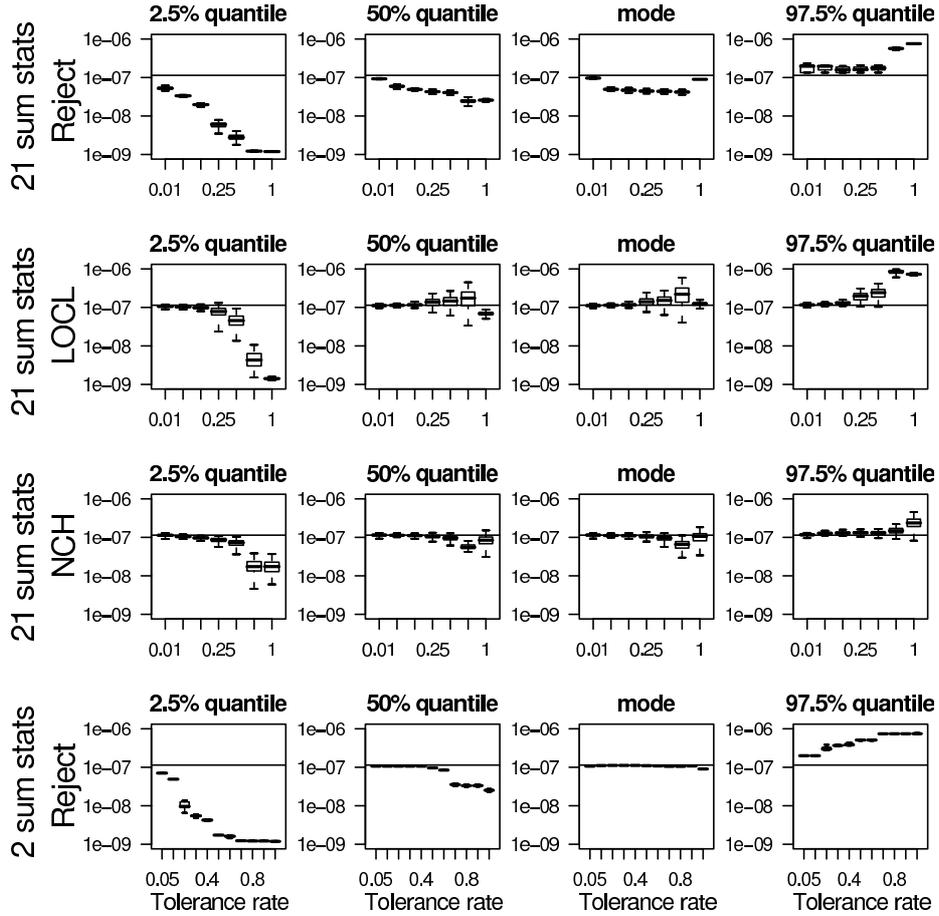}
\end{center}
\caption{Boxplots of the $M=200$ estimated modes and quantiles ($2.5\%$, $50\%$, and $97.5\%$) of the partial posterior distributions of $\lambda_1$. For each ABC method and each value of the tolerance rate, $200$ posterior distributions are computed for each of the 200 synthetic data sets. The horizontal lines correspond to the true value $\lambda_1=1.14 \times 10^{-7}$ used when simulating the 200 synthetic data sets. The different tolerance rates are 0.01, 0.05, 0.10, 0.25, 0.50, 0.50, 0.75, and 1 for all the ABC methods except the rejection scheme with the two summary statistics. For the latter method, the tolerance rates are 0.007, 0.02, 0.06, 0.13, 0.27, 0.37, 0.45, 0.53, 0.66, 0.80, 1.}
\label{fig:boxplots}
\end{figure}

\noindent \textbf{Point estimates of $\theta$ and credibility intervals} Figure \ref{fig:boxplots} displays the boxplots of the 200 estimated modes, medians, $2.5\%$ and $97.5\%$ quantiles of the posterior distribution for $\lambda_1$ as a function of the tolerance rate $P_\delta$ (see Figures 1 and 2 of the supplementary material for $\lambda_2$ and $\lambda_3$).
First, we find that the medians and modes are equivalent except for the rejection method with 21 summary statistics for which the mode is less biased. For the lowest tolerance rates, the point estimates obtained with the four possible methods are close to the value $\lambda_1=1.14\times 10^{-7}$ used in the simulations, with smaller CI for the LOCL and NCH variants. When increasing the tolerance rate, the bias of the point estimates obtained with the rejection method with 21 summary statistics slightly increases. By contrast, up to tolerance rates smaller than $50\%$, the biases of the point estimates obtained with the three other methods remain small. As can be expected, the widths of the CI obtained with the rejection methods increase with the tolerance rate while they remain considerably less variable for the methods with regression adjustment.\\

\begin{figure}[h]
\begin{center}
\includegraphics[height = 9cm]{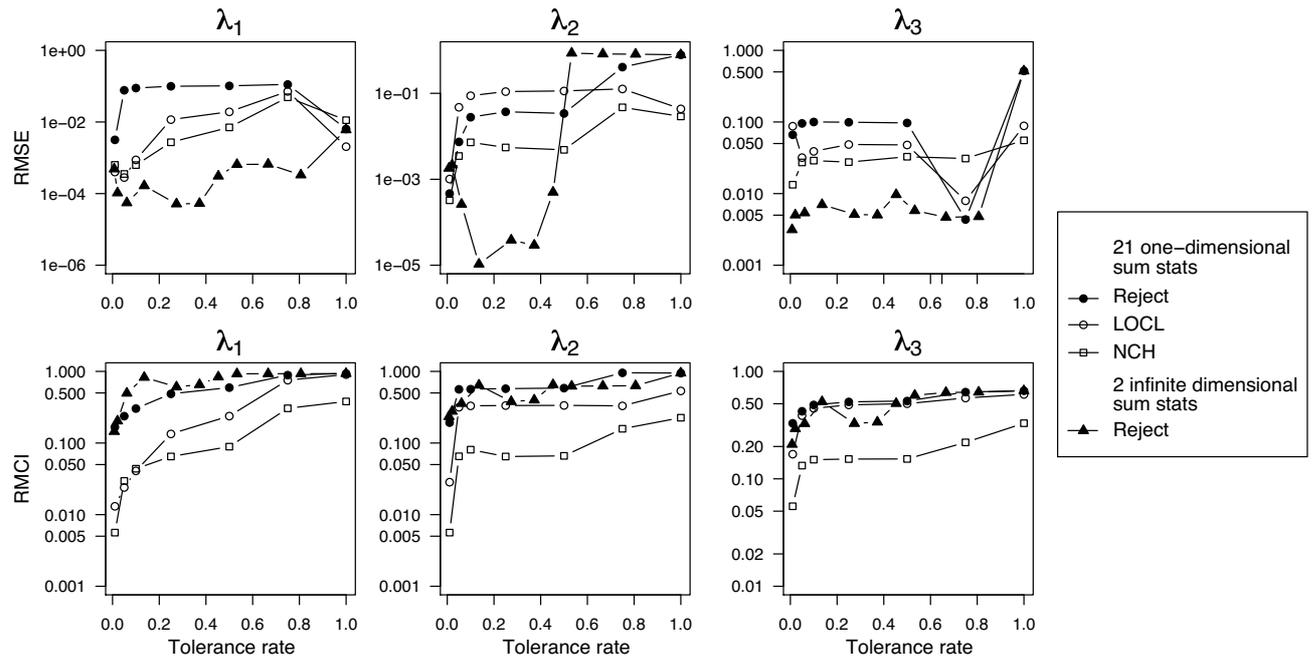}
\end{center}
\caption{Rescaled mean squared error (RMSE) of the posterior mode and rescaled mean credibility interval (RMCI).}
\label{fig:RMSE}
\end{figure}

\noindent \textbf{Mean square error} For further comparison of the different methods, we compute the rescaled mean square errors (RMSEs). The RMSEs are computed on a log scale and rescaled by the range of the prior distribution so that
\begin{equation}
\mbox{RMSE}(\lambda_j)=\frac{1}{M}\sum_{k=1}^{M} \frac{(\log(\hat{\lambda^k_j})-\log(\lambda_j))^2}{\rm{Range}(\rm{prior}(\lambda_j))^2}, \quad j=1,2,3,
\end{equation}
where $\hat{\lambda}_j^k$ is a point estimate obtained with the $k^{\rm{th}}$ synthetic data set. We find that the smallest values of the RMSEs are usually reached for the lowest value of the tolerance rate (see Figure \ref{fig:RMSE}).
For $\lambda_1$ and  $\lambda_2$, the RMSEs of the point estimates obtained with the four different methods are comparable for the lowest tolerance rate. However, the smallest values of the RMSEs are always found when performing the rejection method with the two sufficient summary statistics $R^1$ and $R^2$. This finding is even more pronounced for the parameter $\lambda_3$.\\

\noindent \textbf{Rescaled mean credibility intervals} To compare the whole posterior distributions obtained with the four different methods, we additionally compute the different CIs. The rescaled mean CI (RMCI) is defined as follows
\begin{equation}
\mbox{RMCI}=\frac{1}{M}\sum_{k=1}^{M} \frac{|IC^k_j|}{\rm{Range}(\rm{prior}(\lambda_j))} , \quad j=1,2,3,
\end{equation}
where $|IC^k_j|$ is the length of the $k^{\rm{th}}$ estimated $95\%$ CI for the parameter $\lambda_j$. As displayed by Figure \ref{fig:boxplots}, the CIs obtained with smooth rejection increase importantly with the tolerance rate whereas such an important increase is not observed with regression adjustment. In Figure \ref{fig:RMSE}, the CIs obtained with the NCH method are clearly the thinnest, those obtained with the rejection methods are the widest and those obtained with the LOCL method have intermediate width. In the following, we perform the NCH correction when considering the finite dimensional vector of summary statistics. This choice is motivated by the small RMSEs and RMCIs obtained with the NCH method (Figure \ref{fig:RMSE}).

\section{Application to the Cuban HIV-AIDS epidemic}\label{sectioncuba}

We calibrate the SIR model with contact-tracing to the Cuban HIV-AIDS database. We consider two methods: smooth rejection ABC with the two path-valued summary statistics (Section \ref{sectionABC}), and the NCH-ABC with the vector of summary statistics (Section \ref{sec:correction}). For the Cuban data, this vector is of dimension $51$.

\subsection{Parameter calibration and goodness of fit}\label{sectionparametercalibration}

For the ABC algorithm, we perform a total of 100,000 simulations, we consider the two different rates of contact-tracing detection (\ref{formlambda3}), and we use the same initial conditions and priors as in Section \ref{sectionvalidation}. To set the value of the tolerance rate $P_{\delta}$, we consider the $15$ first years of the epidemic as the training data and choose the value of the tolerance rate $P_{\delta}$ that minimizes the prediction error at the end of the epidemic ($T=21.5$)
\begin{equation}
{\rm Pred \, Error}=
\E \left[\frac{|R^1_{21.5}(P_{\delta})-R^1_{obs,21.5}|}{R^1_{obs,21.5}} + \frac{|R^2_{21.5}(P_{\delta})-R^2_{obs,21.5}|}{R^2_{obs,21.5}}\right].
\end{equation}

\begin{figure}[h]
\begin{center}
\includegraphics[height = 12cm]{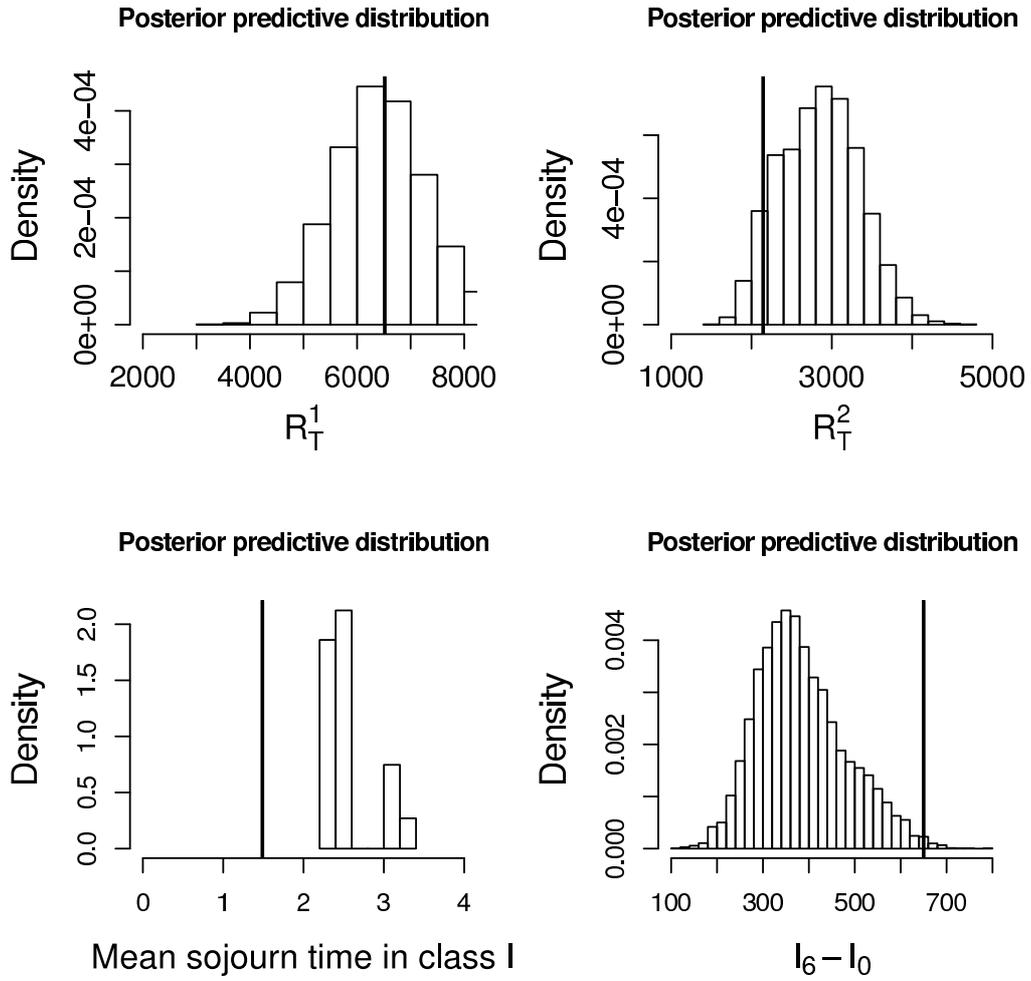}
\end{center}
\caption{Bayesian posterior predictive distributions of $R^1_{21.5}$, $R^2_{21.5}$, $I_6$, and the mean sojourn time in the class I. The SIR model corresponds to the model with frequency dependence for contact-tracing detection. The partial posterior samples are obtained with the smooth rejection ABC algorithm  by making use of the 2 infinite-dimensional summary statistics $R^1$ and $R^2$. Tolerance rate of $P_{\delta_1}=P_{\delta_2}=1\%$ are considered for each summary statistic.}
\label{fig:post_rej_model2}
\end{figure}

For the optimal tolerance rate $P_\delta$, we investigate the goodness of fit of the SIR-type model. By simulating paths of the SIR model associated with parameters $\theta$ sampled from the partial posterior distribution, we check if the model reproduces {\it a posteriori} the observed summary statistics \cite{GelmanMeng}. In Figure \ref{fig:post_rej_model2}, we display the Posterior Predictive Distributions (PPD) of different summary statistics. Figure \ref{fig:post_rej_model2} has been obtained by considering the two sufficient summary statistics $R^1$ and $R^2$, using the optimal tolerance rates of $P_{\delta_1}=P_{\delta_2}=1\%$, and considering the model of frequency dependence (second rate in (\ref{formlambda3})). The cumulated number of detected individuals are contained in the ranges of the PPDs.  By contrast, the mean sojourn time in the class I is not contained in the PPD and the observed number of infectious individuals is in the lower tail of the PPD. An explanation might be that an age-structure has to be taken into account for the infection rate in order to capture the non-Markovian effects (\eg \cite{streftarisgibson}). A model with an increasing infection rate could diminish the mean sojourn time in the class I and increase by compensation the number of infections to maintain the infection pressure constant.\\
When considering the model with a mass action principle (first rate in (\ref{formlambda3})), we observed (see Figure 3 of the supplementary material) that the statistic $R^{1}_{obs,21.5}$ is not contained in the PPD. With a mass action principle, the rate of contact-tracing detection increases linearly with the contribution of the detected individuals, and that is too rapid in comparison with the data.\\
Last, we find that the PPDs obtained with the NCH method have extremely wide supports for both rates of contact-tracing detection (see Figure 4 of the supplementary material). These large PPDs suggest that the summary statistics measuring the detections and the infections may contain conflicting signals, which results in a large variance of the partial posterior distribution.

To provide point estimates and CIs, we consider the model that provides the best fit (model with frequency dependence fitted with the two trajectories $R^1$ and $R^2$). For point estimation, we compute the posterior mode. The estimate of $\lambda_1$ is $5.4 \times 10^{-8}$ ($95\%{\rm CI} \; [3.9 \times 10^{-8};2.3 \times 10^{-7}]$), the estimate of $\lambda_2$ is 0.13 ($95\%{\rm CI} \; [0.007;1.17]$), and the estimate of $\lambda_3$ is $0.19$ ($95\%{\rm CI} \; [0.03;0.82]$). The point estimate of the rate of infection $\lambda_1$ implies that the net rate of infection per infectious individuals $\lambda_1 S$ is equal to $0.32$ ($95\%{\rm CI} \; [0.23;1.37]$). This means that the waiting time before an infectious individual, that has not been detected yet, infects an other individual is $3.1$ years ($95\%{\rm CI} \; [0.72;4.34]$).

\subsection{The dynamic of the Cuban HIV-AIDS epidemic}

\begin{figure}[h]
\begin{center}
\includegraphics[height = 12cm]{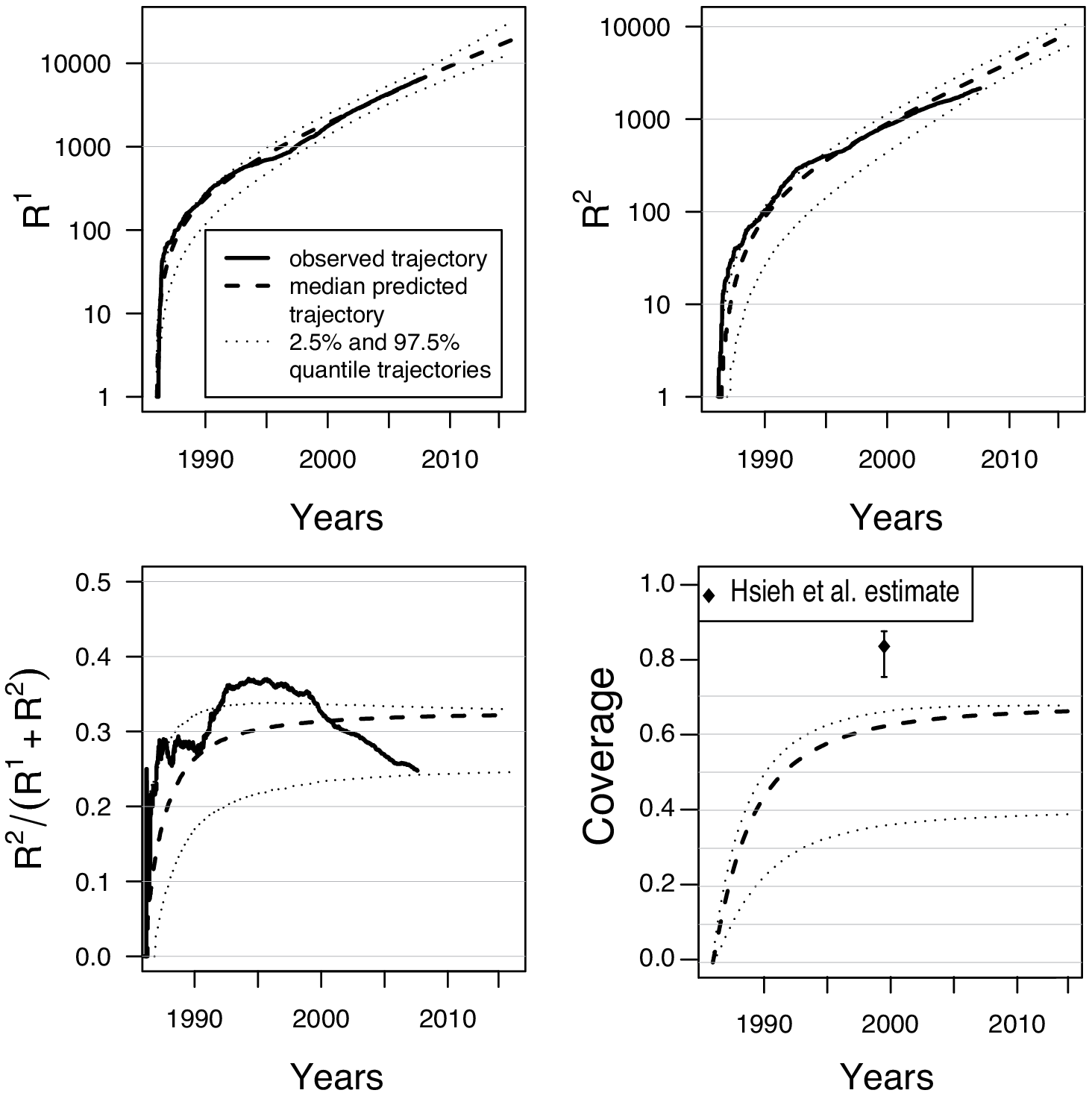}
\end{center}
\caption{Median and $95\%$ credibility intervals of the posterior predictive distributions of $R^1_t$, $R^2_t$, $R^1_t/(R^1_t+R^2_t)$ and coverage $R_t/(I_t+R_t)$ from 1986 to 2015. The coverage is defined as the proportion of known HIV positive individuals. The posterior samples are generated by the rejection scheme with the two summary statistics. A tolerance rate of $P_{\delta}=1\%$ is considered for each summary statistic.}
\label{fig:pred}
\end{figure}

\textbf{Reconstruction of the cumulative numbers of detections} Figure \ref{fig:pred} displays the  dynamics predicted by the SIR model for the numbers of individuals detected by random
screening, contact-tracing and for the number of unknown infectious individuals. Interestingly, there is a really good fit between the real and predicted numbers of individuals detected by random screening except between 1992 and 1995.
This period corresponds to the period of crisis that followed the collapse of the Soviet Union and during which the HIV detection system received less attention. We also find a slight discrepancy in the recent years (2000-2007) between the real and predicted numbers of detections by contact-tracing, which may reveal a weakening in the contact-tracing system.  An explanation is that a new mode of detection, related to contact-tracing but still counted as random detection, has been developed. This new type of detection is promoted by the family doctors who ask to their patients the names of individuals at risk (H. De Arazoza, personal communication).\\

\noindent \textbf{Performance of the contact-tracing system} When testing the performance of contact-tracing, \cite{hsieh} computed the coverage of the epidemic defined as the percentage of infectious individuals that have been detected $(R^1+R^2)/(I+R^1+R^2)$. As displayed by Figure \ref{fig:pred}, the SIR model predicts a coverage of $62\%$ ($95\% \rm{CI} \; [36\%;66\%]$) in 2000 that is much lower than a coverage of $83\%$ ($95\% \rm{CI} \;[75\%;87\%]$) as inferred by \cite{hsieh}. However, since the PPDs of Figure \ref{fig:post_rej_model2} show that the SIR model predicts less infectious individuals than observed, the coverage might still be overestimated and would consequently be even smaller than $62\%$.

Using this estimation of the coverage, we can compare the rates of detection by random screening and contact-tracing per infectious individual. The estimated per capita rate of random screening is $\lambda_2=0.13$. The per capita rate of contact-tracing equals $\lambda_3 \sum_{i \in {R}} \exp(-c(t-T_i))/(I_t+ \sum_{i \in {R}} \exp (-c(t-T_i)))$. Using a zero-order expansion, we find that this rate can be approximated by the product of $\lambda_3$ with the coverage of the epidemic. Hence, the per capita rate of contact-tracing can be estimated as $0.19\times 0.62 \approx 0.12$ that is almost equal to $\lambda_2$.\\

\noindent \textbf{Predictions} Additionally, simulations of the SIR model provide predictions for the evolution of the HIV dynamic in the forthcoming years. The SIR model predicts that in 2015, $42,000$ ($95\% {\rm CI} \;[29,000;107,000] $) individuals will be infected since the beginning of the epidemic in Cuba. Among these infected individuals, a proportion of $45\%$ ($95\% {\rm CI} \; [29\%;46\%] $) will be detected by random screening and a proportion of $21\%$ ($95\% {\rm CI} \; [10\%;22\% ]$) by contact-tracing. As displayed by Figure \ref{fig:pred}, the SIR-type model with contact-tracing predicts that the total proportion of individuals detected by contact-tracing will reach an asymptote of $32\%$ ($95\% {\rm CI} \; [25\%;33\%]$) in 2015. The total number of infected individuals in 2015 corresponds to $27,000$ ($95\% {\rm CI}\; [19,000;80,000]$) new cases of HIV between July 2007 and January 2015. In the same period of time, the SIR model predicts that $12,000$ individuals ($95\% {\rm CI}\; [9,000;24,000]$) will be detected by random screening and $6,000$ individuals ($95\% {\rm CI} \;[4,000;8,000]$) by contact-tracing.

\section{Conclusions}
In the context of temporal epidemiological data, ABC techniques can provide accurate estimates of the parameters of interest such as the infection and detection rates \cite{mckinleyetal,walkeretal}. ABC relies on simulations of the model and can therefore be applied to various epidemiological models.

For partially observed population and missing infectious times, MCMC methods require to reconstruct the unknown data which can be highly computationally intensive for large populations. For instance, \cite{oneillroberts} and \cite{streftarisgibson} considered MCMC algorithms for populations consisting of about 100 individuals whereas the Cuban HIV-AIDS database contains almost 10,000 known HIV positive individuals, which makes the total (known and unknown) number of infectious individuals even larger. When the dimension of the missing data, the infection times here, is both large and unknown, data imputation with MCMC can be computationally very demanding and takes several days on a parallelized system \cite{chisstersinghferguson}.

However, compared with the abundant MCMC literature, the experience of statisticians with ABC is still rather limited. For MCMC algorithms, theoretical convergence results \cite{tierney} as well as practical methods for monitoring numerical convergence are available \cite{gelman96}. Although theoretical results are now available for ABC \cite{blum}, there is no method  for evaluating the two approximations inherent to ABC (Section \ref{sectionmainpple}). Here, we are in a favorable situation since the full summary statistics are sufficient so that the partial and the full posterior are the same. However, for the second approximation, the practice of conditional density estimation in high dimensions remains an issue. When comparing posterior distributions obtained with MCMC and ABC for a standard SIR model, we find the same modes provided that the tolerance rate is small enough. However, even for the smallest tolerance rate, we find that ABC generates posterior distributions with larger tails compared with MCMC. More generally, ABC applications have been restricted to models with moderate number of parameters whereas MCMC applications can involve a very large number of parameters (\cite{pritchard02}). For models with a substantial numbers of parameters, adaptive ABC algorithms that use the simulations to modify the sampling distribution of the parameter $\theta$, might constitute interesting ways to explore for the future of ABC in epidemiology \cite{sissonfantanaka,robertbeaumontmarincornuet,tonietal}.

In this paper, we consider both finite and infinite dimensional summary statistics for ABC. When comparing  ABC with the two different sets of statistics, we find that the point estimates of the parameters $\lambda_1$, $\lambda_2$, $\lambda_3$, with the smallest quadratic errors are obtained with the rejection method based on the infinite-dimensional statistics. However, the $95\%$ CIs obtained with this method are large and critically depend on the tolerance rate. By contrast, regression-based adjustment methods, and the NCH method more particularly, considerably shorten the CIs and are less sensitive to the tolerance rate. Applications of regression-based ABC methods constitute therefore a solution for `stabilizing' the CIs.  However, no ABC with regression adjustment have been developed so far for infinite-dimensional summary statistics.

In the last section of the paper, we calibrate the SIR model to the Cuban HIV-AIDS data. By comparing the posterior predictive distributions of the total number of detections, we find that the model with a frequency-dependent rate of contact-tracing provides the best fit to the data. Using this model, we compare the present-day rates of contact-tracing and random screening. We find that they are almost the same and equal to $0.13$/individual/year. Converting rates of detection to waiting times before detections, we find that the  waiting time before an individual infected today will be detected is equal to $1/(2\times 0.13)\approx 3.8 $ years. At the time of detection, both types of detection are equally probable. Although it suggests that contact-tracing detection contributes importantly to HIV screening in Cuba, we find that the screening might have been largely incomplete. The percentage of undetected individuals among the infectious individuals might have been underestimated \cite{hsieh} and would be of the order of $40\%$.

\section*{Acknowledgments}

The authors are grateful to Pr. H. De Arazoza of the University of La Havana and to Dr. J. Perez of the National Institute of Tropical Diseases in Cuba for granting them access to the HIV-AIDS database. This work has been supported by the ANR Viroscopy (ANR-08-SYSC-016-03), the ANR MANEGE (ANR-09-BLAN-0215-01), and by the Rh\^one-Alpes Institute of Complex Systems (IXXI). The authors thank the anonymous referees for their valuable suggestions.

\section*{supplementary materials}

Supplementary material is available at http://www.biostatistics.oxfordjournals.org

\providecommand{\noopsort}[1]{}\providecommand{\noopsort}[1]{}\providecommand{%
\noopsort}[1]{}\providecommand{\noopsort}[1]{}

\newpage


\clearpage
\newpage

\clearpage
\newpage

\clearpage
\newpage

\clearpage
\newpage

\clearpage
\newpage

\begin{thebibliography}{99}

\bibitem{banksburnscliff}
H.T. Banks, J.A. Burns, and E.M. Cliff.
\newblock Parameter estimation and identification for systems with delays.
\newblock {\em SIAM Journal of Control and Optimization}, 19(6):791--828, 1981.

\bibitem{robertbeaumontmarincornuet}
M.A. Beaumont, J.-M. Marin, J.-M. Cornuet, and C.P. Robert.
\newblock Adaptive approximate bayesian computation.
\newblock {\em Biometrika}, 96(4):983--990, 2009.

\bibitem{beaumontzhangbalding}
M.A. Beaumont, W.~Zhang, and D.J. Balding.
\newblock Approximate {B}ayesian computation in population genetics.
\newblock {\em Genetics}, 162:2025--2035, 2002.

\bibitem{Beckerbritton}
N.G. Becker and T.~Britton.
\newblock Statistical studies of infectious disease incidence.
\newblock {\em Journal of the Royal Statistical Society: Series B},
  61(2):287--307, 1999.

\bibitem{blum}
M.G. Blum.
\newblock {A}pproximate {B}ayesian {C}omputation: a non-parametric perspective.
\newblock {\em Journal of the American Statistical Association}, 2009.
\newblock to appear.

\bibitem{blumfrancois}
M.G.B. Blum and O.~Fran\c{c}ois.
\newblock Non-linear regression models for {A}pproximate {B}ayesian
  {C}omputation.
\newblock {\em Statistics and Computing}, 20:63--73, 2010.

\bibitem{burnscliffdoughty}
J.A. Burns, E.M. Cliff, and S.E. Doughty.
\newblock Sensitivity analysis and parameter estimation for a model of
  {C}hlamydia {T}rachomatis infection.
\newblock {\em Journal of Inverse Ill-Posed Problems}, 15:19--32, 2007.

\bibitem{cauchemcarrat}
S.~Cauchemez, F.~Carrat, C.~Viboud, A.~J. Valleron, and P.~Y. Boelle.
\newblock A {Bayesian MCMC} approach to study transmission of influenza:
  application to household longitudinal data.
\newblock {\em Statistics in Medicine}, 23:3469--3487, 2004.

\bibitem{cauchemezferguson}
S.~Cauchemez and N.M. Ferguson.
\newblock Likelihood-based estimation of continuous-time epidemic models from
  time-series data: application to measles transmission in london.
\newblock {\em Journal of the Royal Society Interface}, 5:885--897, 2008.

\bibitem{arazozaclemencontran}
S.~Cl\'{e}men\c{c}on, V.C. Tran, and H.~De Arazoza.
\newblock A stochastic {SIR} model with contact-tracing: large population
  limits and statistical inference.
\newblock {\em Journal of Biological Dynamics}, 2(4):392--414, 2008.

\bibitem{auvert}
H.~de~Arazoza, J.~Joanes, R.~Lounes, C.~Legeai, S.~Cl\'emen\c{c}on, J.~Perez,
  and B.~Auvert.
\newblock The {HIV/AIDS} epidemic in {C}uba: description and tentative
  explanation of its low prevalence.
\newblock {\em BMC Infectious Disease}, 7:130, 2007.

\bibitem{arazozacubana}
H.~de~Arazoza, R.~Lounes, J.~Perez, and T.~Hong.
\newblock What percentage of the cuban {HIV-AIDS} epidemic is known.
\newblock {\em Rev Cubana Med Trop}, 55:30--37, 2003.

\bibitem{fan1}
J.~Fan.
\newblock Design-adaptive nonparametric regression.
\newblock {\em Journal of the American Statistical Association},
  87(420):998--1004, 1992.

\bibitem{finken}
B.F Finkenst\"adt, O.N. Bj{\o}rnstad, and B.T. Grenfell.
\newblock A stochastic model for extinction and recurrence of epidemics:
  estimation and inference for measles outbreaks.
\newblock {\em Biostatistics}, 3:493--510, 2002.

\bibitem{gelman96}
A.~Gelman.
\newblock Inference and monitoring convergence.
\newblock In W.R. Gilks, S.~Richardson, and D.J. Spiegelhalter, editors, {\em
  {M}arkov chain {M}onte {C}arlo in practice}, pages 131--144. Chapman \& Hall,
  1996.

\bibitem{GelmanMeng}
A.~Gelman and X-L Meng.
\newblock Model checking and model improvment.
\newblock In W.R. Gilks, S.~Richardson, and D.J. Spiegelhalter, editors, {\em
  {M}arkov chain {M}onte {C}arlo in practice}. Chapman \& Hall, 1996.

\bibitem{GilksRoberts}
W.R. Gilks and G.O. Roberts.
\newblock Strategies for improving {MCMC}.
\newblock In W.R. Gilks, S.~Richardson, and D.J. Spiegelhalter, editors, {\em
  {M}arkov chain {M}onte {C}arlo in practice}. Chapman \& Hall, 1996.

\bibitem{hsieh}
Y.H. Hsieh, H.~de~Arazoza, S.M. Lee, and C.W. Chen.
\newblock Estimating the number of {C}ubans infected sexually by human
  immunodeficiency virus using contact tracing data.
\newblock {\em Int. J. Epidemiol.}, 31(3):679--83, 2002.

\bibitem{mckinleyetal}
T.~McKinley, A.~R. Cook, and R.~Deardon.
\newblock Inference in epidemic models without likelihoods.
\newblock {\em The International Journal of Biostatistics}, 5(1), 2009.

\bibitem{oneill}
P.D. O'Neill.
\newblock A tutorial introduction to {B}ayesian inference for stochastic
  epidemic models using {M}arkov chain {M}onte {C}arlo methods.
\newblock {\em Mathematical Biosciences}, 180:103--114, 2002.

\bibitem{oneillroberts}
P.D. O'Neill and G.O. Roberts.
\newblock Bayesian inference for partially observed stochastic epidemics.
\newblock {\em Journal of the Royal Statistical Society: Series A},
  162:121--129, 1999.

\bibitem{pritchard02}
J.~K. Pritchard, M.~Stephens, and P.~Donnelly.
\newblock Inference of population structure using multilocus genotype data.
\newblock {\em Genetics}, 155:945--959, 2002.

\bibitem{sissonfantanaka}
S.A. Sisson, Y.~Fan, and M.~Tanaka.
\newblock Sequential {M}onte {C}arlo without likelihoods.
\newblock {\em Proc. Nat. Acad. Sci. USA}, 104:1760--1765, 2007.

\bibitem{chisstersinghferguson}
I.~Chis Ster, B.K. Singh, and N.M. Ferguson.
\newblock Epidemiological inference for partially observed epidemics: the
  example of the 2001 foot and mouth epidemic in {G}reat {B}ritain.
\newblock {\em Epidemics}, 1:21--34, 2009.

\bibitem{streftarisgibson}
G.~Streftaris and G.J. Gibson.
\newblock Bayesian inference for stochastic epidemics in closed population.
\newblock {\em Statistical Modelling}, 4(1):63--75, 2004.

\bibitem{tierney}
L.~Tierney.
\newblock Markov chains for exploring posterior distributions.
\newblock {\em Annals of Statistics}, 22(4):1701--1728, 1994.

\bibitem{tonietal}
T.~Toni, D.~Welch, N.~Strelkowa, A.~Ipsen, and M.P. Stumpf.
\newblock Approximate {B}ayesian computation scheme for parameter inference and
  model selection in dynamical systems.
\newblock {\em Journal of The Royal Society Interface}, 6:187--202, 2009.

\bibitem{walkeretal}
D.M. Walker, D.~Allingham, H.W.J. Lee, and M.~Small.
\newblock Parameter inference in small world network disease models with
  {A}pproximate {B}ayesian {C}omputational methods.
\newblock {\em Physica A}, 389:540--548, 2010.

\end{thebibliography}
\end{document}